\documentclass[runningheads,a4paper]{llncs}

\usepackage{amssymb}
\usepackage{graphicx}
\usepackage{wrapfig}
\usepackage{subfloat}
\usepackage{pifont}
\usepackage{float}
\usepackage{url}
\usepackage{tabularx}
\newcolumntype{C}[1]{>{\centering\arraybackslash}p{#1}} 
\newcolumntype{R}[1]{>{\raggedleft\arraybackslash}p{#1}} 

\newcommand{\saucy}{{\bf saucy}}
\newcommand{\nauty}{{\bf nauty}}
\newcommand{\bliss}{{\bf bliss}}
\newcommand{\nishe}{{\bf nishe}}

\begin{document}
\mainmatter  

\title{Conflict Anticipation in the Search for Graph Automorphisms}

\titlerunning{Conflict Anticipation in the Search for Graph Automorphisms}

%
%
\author{Hadi Katebi \and Karem A. Sakallah \and Igor L. Markov}
%

\institute{EECS Department, University of Michigan\\
\{hadik,karem,imarkov\}@umich.edu}

%
%

\maketitle

\begin{abstract}
Effective search for graph automorphisms allows identifying symmetries in many discrete structures, ranging from chemical molecules to microprocessor circuits. Using this type of structure can enhance visualization as well as speed up computational optimization and verification. Competitive algorithms for the graph automorphism problem are based on efficient partition refinement augmented with group-theoretic pruning techniques. In this paper, we improve prior algorithms for the graph automorphism problem by introducing \emph{simultaneous refinement of multiple partitions}, which enables the anticipation of future conflicts in search and leads to significant pruning, reducing overall runtimes. Empirically, we observe an exponential speedup for the family of Miyazaki graphs, which have been shown to impede leading graph-automorphism algorithms.
\end{abstract}

\section{Introduction}
\label{sec:intro}
 
An \emph{automorphism} (\emph{symmetry}) of a graph is a \emph{permutation} of the graph's vertices that preserves the graph's edge relation. The set of all symmetries of a graph forms a \emph{group}\footnote{A group is an algebraic structure comprising a non-empty set of elements with a binary operation that is \emph{associative}, admits an \emph{identity} element, and is \emph{invertible}. For example, the set of integers with addition forms a group. A \emph{generating set} of a group is a subset of the group's elements whose combinations under the group operation generate the entire group.} under functional composition. The \emph{graph automorphism problem} seeks a generating set for the automorphism group of a graph. Closely related to graph automorphism is the problem of \emph{canonical labeling} which assigns a unique signature to a graph that is invariant under all possible labelings of its vertices. Graph automorphisms and canonical labelings are related to the functional properties of the combinatorial objects in question. In a representative application developed in \cite{Aloul-DAC02,Aloul-DAC03}, a CNF (conjunctive normal form) formula is modeled by a graph and passed to a symmetry detection program. During subsequent symmetry-breaking, these symmetries are used to augment the formula with a set of symmetry-breaking predicates. These predicates do not change the formula's satisfiability, but help SAT solvers prune away symmetric portions of the search space.

Graph symmetry and canonical labeling have been extensively studied over the past five decades. The \nauty{} program \cite{nautyUG,McKay81}, developed by McKay in 1981, pioneered the first high-performance algorithms that inspired all subsequent tools. Almost two decades later, Darga et al \cite{Darga-DAC04} observed that the use of an adjacency matrix in \nauty{} could lead to asymptotic inefficiencies in dealing with sparse graphs. This motivated the development of a new tool called \saucy{} \cite{Darga-DAC04,Darga-DAC08,Katebi-2010}, which was limited to just finding a set of symmetry generators, but was three orders of magnitude faster than \nauty{} on very large and very sparse graphs. Closely following \nauty{}'s canonical labeling algorithms were two other tools, namely, \bliss{} \cite{Junttila07,Junttila-2011} and \nishe{} \cite{Tener08}. The search routines in \bliss{} improved the handling of large and sparse graphs, and the branching heuristics in \nishe{} facilitated a polynomial-time solution for the Miyazaki graphs \cite{Miyazaki_1997}, a family of graphs that \nauty{} requires exponential time to process. 

Since the emergence of the first version of \saucy{} in 2004 (\saucy{} 1.1) \cite{Darga-DAC04}, different algorithmic enhancements improved \saucy{}'s performance over a wide range of graphs with both theoretical and practical interest. The second version of \saucy{} (\saucy{} 2.0) \cite{Darga-DAC08} incorporated the observation that the symmetry generators of sparse graphs were mostly sparse. The major algorithmic changes that were introduced in \saucy{} 2.0 separated the search for symmetries from the search for a canonical labeling. Further improvements to \saucy{}'s data structures and algorithms were reported in \saucy{} 2.1 \cite{Katebi-2010}.

In this paper, we present \saucy{} 3.0 which performs \emph{simultaneous partition refinement} to anticipate and avoid possible future conflicts. The procedure augments the method introduced in \saucy{} 2.1 whereby nodes in the search tree represent sets of vertex permutations encoded by an ordered partition pair (OPP) of graph vertices. The basic idea of the new procedure is to refine the top and bottom partitions of an OPP at the same time, making sure that the two partitions conform to each other (according to the graph's edge relation) after each refinement step. We implemented this enhancement in \saucy{} 3.0 and tested its performance on a wide variety of graph benchmarks. Our experimental evaluation shows that this enhancement can significantly prune the search tree for many graph families, such as the Miyazaki graphs. Furthermore, the concept of simultaneous refinement helps us better understand and explain the validity of some of the algorithms that were previously presented in \saucy{} 2.1.

In the remainder, we first review some preliminaries in Section \ref{sec:prelim}. Then, we discuss \saucy{}'s baseline algorithms in Section \ref{sec:algorithms}. The new partitioning algorithm based on the concept of simultaneous refinement is presented in Section \ref{sec:refine}. Section \ref{sec:matching} establishes the correctness of ``matching OPP'' pruning (this pruning mechanism was presented in \saucy{} 2.1). The results of our experimental study are provided in Section \ref{sec:results}. Finally, we discuss conclusions in Section \ref{sec:conc}. 

\section{Preliminaries}
\label{sec:prelim}

We assume familiarity with basic notions from group theory, including such concepts as groups, subgroups, group generators, cosets, orbit partition, etc. Information on different group theoretic concepts is available in many abstract algebra texts such as \cite{Fraleigh00}. In this paper, we focus on the automorphisms of an $n$-vertex \emph{colored graph} $G$ whose vertex is $V=\{0,1,...,n-1\}$. A \emph{permutation} of $V$ is a bijection from $V$ to $V$, and a \emph{symmetry} of $G$ is a permutation of $V$ that preserves $G$'s edge relation. Permutation $\alpha$, when applied to $G$, produces the permuted graph $G^\alpha$. Every graph has a trivial symmetry, called the \emph{identity}, that maps each vertex to itself. The set of symmetries of $G$ forms a \emph{group} under functional composition. This group is the \emph{symmetry group} of $G$, and is denoted by $Aut(G)$. Given $G$, the objective of any symmetry detection tool is to find a set of \emph{group generators} for $Aut(G)$.

An \emph{ordered partition} $\pi=[W_1 \vert W_2 \vert \cdots \vert W_m]$ of $V$ is an ordered list of non-empty pair-wise disjoint subsets of $V$ whose union is $V$. The subsets $W_i$ are called the \emph{cells} of the partition. Ordered partition $\pi$ is \emph{unit} if $m=1$ (i.e., $W_1 = V$) and \emph{discrete} if $m=n$ (i.e., $|W_i|=1$ for $i=1,\cdots,n$). An \emph{ordered partition pair (OPP)} $\pi$ is specified as 
\[
\Pi  = \left[ {\begin{array}{*{20}c}
   {\pi _T }  \\
   {\pi _B }  \\
\end{array}} \right] = \left[ {\begin{array}{*{20}c}
   {T_1 \left| {T_2 \left| { \cdots \left| {T_m } \right.} \right.} \right.} \hfill  \\
   {B_1 \left| {B_2 \left| { \cdots \left| {B_k } \right.} \right.} \right.} \hfill  \\
\end{array}} \right]
\]
with $\pi_T$ and $\pi_B$ referred to, respectively, as the top and bottom ordered partitions of $\pi$. OPP $\pi$ is \emph{isomorphic} if $m=k$ and $|T_i|=|B_i|$ for $i=1,\cdots,m$; otherwise it is \emph{non-isomorphic}. In other words, an OPP is isomorphic if its top and bottom partitions have the same number of cells, and corresponding cells have the same cardinality. An isomorphic OPP is \emph{matching} if its corresponding non-singleton cells are \emph{identical}. We will refer to an OPP as discrete (resp. unit) if its top and bottom partitions are discrete (resp. unit). 

OPPs lie at the heart of \saucy{}'s symmetry detection algorithms, since each OPP compactly represents a set of permutations. This set of permutations might be empty (non-isomorphic OPP), might have only one permutation (discrete OPP), or might consist of up to $n!$ permutations (unit OPP). Several OPP examples and the permutation set encoded by them are provided below.
\begin{itemize}
\item Discrete OPP:
$
\left[ {\left. {\begin{array}{*{20}c}
   2  \\
   1  \\
\end{array}} \right|\left. {\begin{array}{*{20}c}
   0  \\
   2  \\
\end{array}} \right|\begin{array}{*{20}c}
   1  \\
   0  \\
\end{array}} \right] = \left\{ {\left( {0\;2\;1} \right)} \right\}
$

\item Unit OPP: 
$
\left[ {\begin{array}{*{20}c}
   {0,1,2} \hfill  \\
   {0,1,2} \hfill  \\
\end{array}} \right] = \left\{ {\iota ,\left( {0\;1} \right),\left( {0\;2} \right),\left( {1\;2} \right),\left( {0\;1\;2} \right),\left( {0\;2\;1} \right)} \right\}
$

\item Isomorphic OPP: 
$
\left[ {\left. {\begin{array}{*{20}c}
   2  \\
   1  \\
\end{array}} \right|\begin{array}{*{20}c}
   {0,1}  \\
   {2,0}  \\
\end{array}} \right] = \left\{ {\left( {1\;2} \right),\left( {0\;2\;1} \right)} \right\}
$

\item Matching OPP:$
\left[ {\left. {\begin{array}{*{20}c}
   1  \\
   3  \\
\end{array}} \right|\left. {\begin{array}{*{20}c}
   {0,2,4}  \\
   {0,2,4}  \\
\end{array}} \right|\begin{array}{*{20}c}
   3  \\
   1  \\
\end{array}} \right] = \left( {1\;3} \right) \circ S_3 \left( {\left\{ {0,2,4} \right\}} \right)
$

\item Non-isomorphic OPPs:
$
\left[ {\begin{array}{*{20}c}
   {\left. {0,2} \right|1} \hfill  \\
   {\left. 1 \right|2,0} \hfill  \\
\end{array}} \right] = \emptyset ,\left[ {\begin{array}{*{20}c}
   {\left. {\left. 2 \right|0} \right|1} \hfill  \\
   {\left. 1 \right|2,0} \hfill  \\
\end{array}} \right] = \emptyset 
$
\end{itemize}

\section{Baseline Algorithms}
\label{sec:algorithms}

Similar to other combinatorial search algorithms, \saucy{} explores the space of permutations by building a search tree and systematically traversing it. However, the representation of search nodes as OPPs in \saucy{} is unique. The root of the tree is a unit OPP which is initially \emph{refined} based on the colors and degrees of the vertices of the input graph. The depth-first traversal of the permutation space is started by choosing a \emph{target} vertex from a non-singleton cell of the top partition and mapping it to all the vertices of the corresponding cell of the bottom partition. To propagate the \emph{constraints of the graph} (i.e. the graph's edge relation), partition refinement is invoked after each mapping decision. The mapping procedure continues until the OPP becomes discrete, matching, or non-isomorphic (the latter is referred to as a \emph{conflict}). In either case, \saucy{} backtracks one level up, and maps the target vertex to the remaining candidate vertices. The search ends when all possible mappings are exhausted. 

In addition to partition refinement, \saucy{} exploits two types of pruning mechanisms: \emph{group-theoretical} and \emph{OPP-based}. To enable group-theoretical pruning, namely \emph{coset} and \emph{orbit} pruning, the left-most path of the tree should correspond to a sequence of \emph{subgroup stabilizers} ending in the identity. In other words, the decisions along the left-most path maps each vertex to itself. This phase of the search is called \emph{subgroup decomposition}. Note that no such requirement is needed in the remaining parts of the search tree. 
In contrast, OPP-based pruning mechanisms are optional techniques that assist \saucy{}'s algorithms to avoid unnecessary search. Two of these techniques, embedded in \saucy{} 2.1, are \emph{non-isomorphic OPP} and \emph{matching OPP} pruning. 

In this paper, we introduce an enhanced partition refinement procedure that refines the top and bottom partitions of an OPP simultaneously. Our simultaneous refinement anticipates the conflicts that might arise in certain subtrees, and prunes the entire subtree without exploring it. The idea here is to capture conflicts that might be overlooked by the conventional refinement procedure.

\section{Conflict Anticipation via Simultaneous Refinement}
\label{sec:refine}

Partition refinement in \saucy{} is adapted from \nauty{}, and \nauty{}'s refinement is based on the concept of \emph{equitable} partitions. Partition $\pi=[W_1 \vert W_2 \vert \cdots \vert W_m]$ is equitable (with respect to graph G) if, for all $v_1, v_2 \in W_i$ ($1\le i\le m$), the number of neighbors of $v_1$ in $W_j$ ($1\le j\le m$) is equal to the number of neighbors of $v_2$ in $W_j$. Although \saucy{}'s partition refinement is adapted from \nauty{}, the search tree in \saucy{} is completely different from that in \nauty{}. The nodes of \nauty{}'s tree are {\bf single ordered partitions}, while the nodes of \saucy{}'s tree are {\bf ordered partition pairs}. In \nauty{}, an equitable partition is obtained by invoking partition refinement after each vertex \emph{individualization}. Extending this to OPPs, the refinement procedure in \saucy{} refines both partitions of an OPP \emph{simultaneously} after each mapping decision, until 1) both partitions become equitable and the resulting OPP is isomorphic, or 2) the resulting OPP is non-isomorphic indicating an empty set of permutations, i.e., a conflict. In \saucy{} 2.1 and earlier, simultaneous refinement was basically an algorithmic enhancement that detected conflicts (if any existed) earlier during refinement, without fully establishing an equitable OPP (an OPP whose top and bottom partitions are both equitable), and then examining the resulting OPP to see whether it was isomorphic/non-isomorphic. In implementation, \saucy{} first refines the top partition until it becomes equitable, records where the cell splits occur, then starts refining the bottom partition, and compares the splitting locations of the bottom to the top whenever a new split occurs (i.e., checks the isomorphism of the two partitions after each split).  

In this section, we argue that the significance of simultaneous refinement is not limited to the early detection of ``non-isomorphic equitable OPPs''. In particular, we demonstrate cases where the resulting equitable OPP is isomorphic, but the OPP still violates the edge relation of the graph. We illustrate such a case, and explain why conventional refinement fails to detect the conflict in that case. We then present an \emph{enhanced simultaneous refinement} procedure that detects such cases and does not explore them. We discuss the impact of our proposed refinement procedure on the search tree constructed for our example. 

\begin{figure}[t]
\centering
\includegraphics[scale=0.65,keepaspectratio=true]{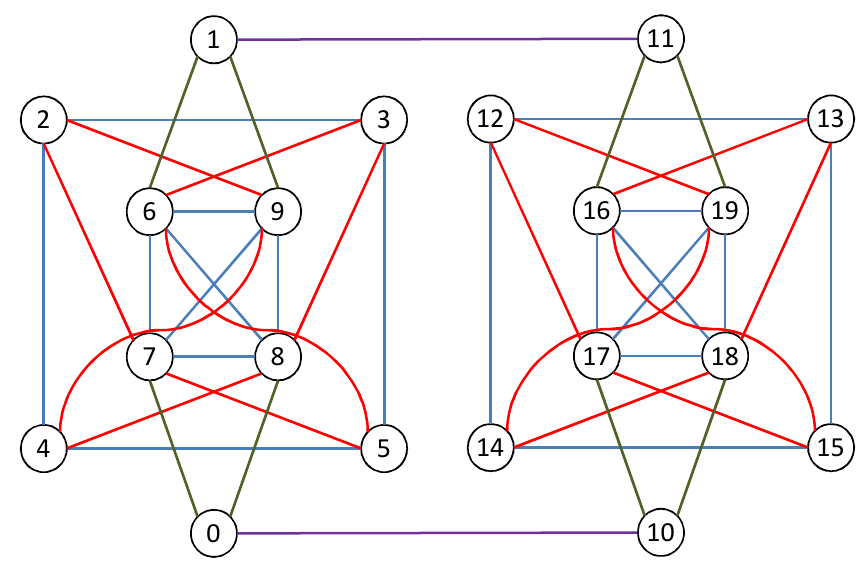}
\caption{A 20-vertex 46-edge graph with symmetry group of size 32.}
\label{fig:graph}
\end{figure}

\begin{figure}[t]
\centering
\includegraphics[scale=0.5,keepaspectratio=true]{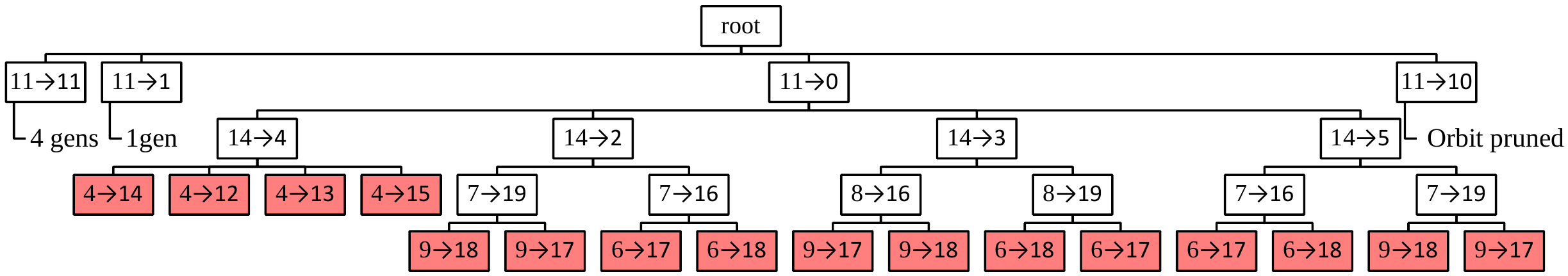}
\caption{The search tree constructed by \saucy{} 2.1 for the graph in Figure \ref{fig:graph}.}
\label{fig:tree1}
\end{figure}

Consider the 20-vertex 46-edge graph shown in Figure \ref{fig:graph}. The search tree generated by \saucy{} 2.1 for this graph is shown in Figure \ref{fig:tree1}. This search tree produces $16$ conflicts (non-isomorphic OPPs), indicated by red-shaded  nodes. In the remainder of this section, we focus on the path from the root that maps $11\mapsto 0$ and then $14\mapsto 4$. The OPPs in Figure \ref{fig:OPPs}, labeled with (\ref{eqn:OPP1}), (\ref{eqn:OPP2}) and (\ref{eqn:OPP3}), represent the nodes of the search tree at the root, after mapping $11\mapsto 0$, and after mapping $14\mapsto 4$, respectively.

\begin{subfigures}
\label{fig:nodes}
\begin{figure}[t]
\begin{equation}
\label{eqn:OPP1}
\left[ {\left. {\begin{array}{*{20}c}
   {11,10,1,0}   \\
   {11,10,1,0} \\
\end{array}} \right|\left. {\begin{array}{*{20}c}
   {15,12,14,13,5,2,4,3}  \\
   {15,12,14,13,5,2,4,3}  \\
\end{array}} \right| \begin{array}{*{20}c}
   {18,19,17,16,8,9,7,6}  \\
   {18,19,17,16,8,9,7,6}  \\
\end{array}} \right]
\end{equation}
\begin{equation}
\label{eqn:OPP2}
\left[ {\left. {\begin{array}{*{20}c}
   0   \\
   11 \\
\end{array}} \right|\left. \hspace{-3pt} {\begin{array}{*{20}c}
   10  \\
   1  \\
\end{array}} \right|\left. \hspace{-3pt} {\begin{array}{*{20}c}
   1  \\
   10  \\
\end{array}} \right|\left. \hspace{-3pt} {\begin{array}{*{20}c}
   11  \\
   0  \\  
\end{array}} \right|\left. \hspace{-3pt} {\begin{array}{*{20}c}
   {14,12,13,15}  \\
   {4,3,5,2}  \\
\end{array}} \right|\left. \hspace{-3pt} {\begin{array}{*{20}c}
   {2,4,5,3}  \\
   {13,14,12,15}  \\
\end{array}} \right|\left. \hspace{-3pt} {\begin{array}{*{20}c}
   {17,18}  \\
   {9,6}  \\
\end{array}} \right|\left. \hspace{-3pt} {\begin{array}{*{20}c}
   {8,7}  \\
   {19,16}  \\
\end{array}} \right|\left. \hspace{-3pt} {\begin{array}{*{20}c}
   {6,9}  \\
   {18,17}  \\
\end{array}} \right| \begin{array}{*{20}c}
   {16,19}  \\
   {7,8}  \\
\end{array}} \right]
\end{equation}
\begin{equation}
\label{eqn:OPP3}
\left[ {\left. {\begin{array}{*{20}c}
   0   \\
   11 \\
\end{array}} \right|\left. \hspace{-3pt} {\begin{array}{*{20}c}
   10  \\
   1  \\
\end{array}} \right|\left. \hspace{-3pt} {\begin{array}{*{20}c}
   1  \\
   10  \\
\end{array}} \right|\left. \hspace{-3pt} {\begin{array}{*{20}c}
   11  \\
   0  \\
\end{array}} \right|\left. \hspace{-3pt} {\begin{array}{*{20}c}
   13  \\
   3  \\
\end{array}} \right|\left. \hspace{-3pt} {\begin{array}{*{20}c}
   12  \\
   2  \\
\end{array}} \right|\left. \hspace{-3pt} {\begin{array}{*{20}c}
   15  \\
   5  \\
\end{array}} \right|\left. \hspace{-3pt} {\begin{array}{*{20}c}
   14  \\
   4  \\
\end{array}} \right|\left. \hspace{-3pt} {\begin{array}{*{20}c}
   {2,4,5,3}  \\
   {13,14,12,15}  \\
\end{array}} \right|\left. \hspace{-3pt} {\begin{array}{*{20}c}
   17  \\
   6  \\
\end{array}} \right|\left. \hspace{-3pt} {\begin{array}{*{20}c}
   18  \\
   9  \\
\end{array}} \right|\left. \hspace{-3pt} {\begin{array}{*{20}c}
   {8,7}  \\
   {19,16}  \\
\end{array}} \right|\left. \hspace{-3pt} {\begin{array}{*{20}c}
   {6,9}  \\
   {18,17}  \\
\end{array}} \right|\left. \hspace{-3pt} {\begin{array}{*{20}c}
   16  \\
   7  \\
\end{array}} \right| \begin{array}{*{20}c}
   19  \\
   8  \\
\end{array}} \right]
\end{equation}
\caption{The search nodes of the tree in Figure \ref{fig:tree1}. OPP (\ref{eqn:OPP1}) is at the root, OPP (\ref{eqn:OPP2}) is after mapping $11\mapsto 0$, and OPP (\ref{eqn:OPP3}) is after mapping $14\mapsto 4$.}
\label{fig:OPPs}
\end{figure}
\begin{figure}[!ht]
\begin{equation} \label{eqn:top1} 
\left[ \begin{tabular*}{0.6\textwidth}{@{\extracolsep{\fill}}c|c|c|c|c|c|c|c|c|c|c} 0 & 10 & 1 & 11 & 12,13,15 & 14 & 2,4,5,3 & 17,18 & 8,7 & 6,9 & 16,19 \end{tabular*} \right]
\end{equation}
\begin{equation} \label{eqn:top2} 
\left[ \begin{tabular*}{0.6\textwidth}{@{\extracolsep{\fill}}c|c|c|c|c|c|c|c|c|c|c|c|c|c} 0 & 10 & 1 & 11 & 13 & 12,15 & 14 & 2,4,5,3 & 18 & 17 & 8,7 & 6,9 & 16 & 19 \end{tabular*} \right]
\end{equation}
\begin{equation} \label{eqn:top3} 
\left[ \begin{tabular*}{0.6\textwidth}{@{\extracolsep{\fill}}c|c|c|c|c|c|c|c|c|c|c|c|c|c|c} 0 & 10 & 1 & 11 & 13 & 12 & 15 & 14 & 2,4,5,3 & 18 & 17 & 8,7 & 6,9 & 16 & 19 \end{tabular*} \right]
\end{equation}
\caption{The refinement of the top partition of OPP (\ref{eqn:OPP2}) to get OPP (\ref{eqn:OPP3}).}
\label{fig:top}
\end{figure}
\begin{figure}[!ht]
\begin{equation} \label{eqn:bot1} 
\left[ \begin{tabular*}{0.6\textwidth}{@{\extracolsep{\fill}}c|c|c|c|c|c|c|c|c|c|c} 11 & 1 & 10 & 0 & 3,5,2 & 4 & 13,14,12,15 & 9,6 & 19,16 & 18,17 & 7,8 \end{tabular*} \right]
\end{equation}
\begin{equation} \label{eqn:bot2} 
\left[ \begin{tabular*}{0.6\textwidth}{@{\extracolsep{\fill}}c|c|c|c|c|c|c|c|c|c|c|c|c|c} 11 & 1 & 10 & 0 & 3 & 5,2 & 4 & 13,14,12,15 & 9 & 6 & 19,16 & 18,17 & 7 & 8 \end{tabular*} \right]
\end{equation}
\begin{equation} \label{eqn:bot3} 
\left[ \begin{tabular*}{0.6\textwidth}{@{\extracolsep{\fill}}c|c|c|c|c|c|c|c|c|c|c|c|c|c|c} 11 & 1 & 10 & 0 & 3 & 2 & 5 & 4 & 13,14,12,15 & 9 & 6 & 19,16 & 18,17 & 7 & 8 \end{tabular*} \right]
\end{equation}
\caption{The refinement of the bottom partition of OPP (\ref{eqn:OPP2}) to get OPP (\ref{eqn:OPP3}).}
\label{fig:bot}
\end{figure}
\end{subfigures}

In \saucy{} 2.1, the isomorphic OPP (\ref{eqn:OPP3}), obtained after mapping $14\mapsto 4$, is not considered to be a conflict node and triggers further vertex mappings (namely, $4\mapsto 14$, $4\mapsto 12$, $4\mapsto 13$, and $4\mapsto 15$). However, this OPP violates the edge relation of the graph in Figure \ref{fig:graph}. To see this, consider the edge that connects $13$ to $16$. This edge, according to OPP (\ref{eqn:OPP3}), should be mapped to another edge that connects $3$ to $7$, since OPP (\ref{eqn:OPP3}) maps $13\mapsto 3$, and $16\mapsto 7$. Nevertheless, no such edge exists between 3 and 7 in Figure \ref{fig:graph}, and hence, OPP (\ref{eqn:OPP3}) is a conflict.

The question now is why the refinement procedure failed to detect the above conflict? Or, in other words, why was OPP (\ref{eqn:OPP3}) found to be isomorphic? To answer this question, we should follow the trace of the refinement procedure which is performed on OPP (\ref{eqn:OPP2}) to get OPP (\ref{eqn:OPP3}) after mapping $14\mapsto 4$. As elaborated earlier, \saucy{} first refines the top partition until it becomes equitable, then refines the bottom partition and checks the isomorphism of the bottom to the top whenever a new split occurs. The step by step refinement of the top and bottom partitions when $14\mapsto 4$ is shown in Figure \ref{fig:top} and Figure \ref{fig:bot}, respectively. 

The refinement on the top starts by first making 14 a singleton cell (partition (\ref{eqn:top1})). According to the graph of Figure \ref{fig:graph}, 14 is connected to 12,15,18 and 19, but not to 13, 17 and 16. Hence, refinement separates 12 and 15 from 13 (this makes 13 a singleton cell), 18 from 17, and 19 from 16 (partition (\ref{eqn:top2})). The refinement continues by looking at the connections of one of the newly created cells. Here, \saucy{} picks the singleton cell 16. According to the graph, 16 is connected to 11,13,15,17,18 and 19. This separates 15 from 12 (partition (\ref{eqn:top3})). The top partition is now equitable, i.e., no further refinement is implied.

After refining the top partition, \saucy{} starts refining the bottom partition. This is done by first making 4 a singleton cell (partition (\ref{eqn:bot1})). Since 4 is connected to 2,5,8 and 9, refinement separates 2 and 5 from 3 (this makes 3 a singleton cell), 9 from 6, and 8 from 7 (partition (\ref{eqn:bot2})). Note that, at this point, partition (\ref{eqn:bot2}) is isomorphic to partition (\ref{eqn:top2}), i.e., no conflict is detected. This time \saucy{} picks the singleton cell 7, since it had previously chosen 16 from the top, and 7 is at the same index on the bottom as 16 on the top. According to the graph, 7 is connected to 0,2,5,6,8 and 9. Since 7 is connected to both 2 and 5, no further refinement is implied. At this point, \saucy{} should \emph{detect the conflict that 16 on the top separated 15 from 12, but 7 on the bottom did not distinguish 2 from 5}. However, since no new cell is created on the bottom, \saucy{} does not invoke the isomorphism check, and falsely assumes that the bottom stays isomorphic to the top. Note that the failure to detect this conflict is not a bug in refinement, since \nauty{}'s (and essentially \saucy{}'s) refinement procedure refines one partition at a time, and checks isomorphism once both partitions are equitable. After refining based on 7, \saucy{} refines based on 6. Vertex 6 is connected to 1,3,5,7,8 and 9. Since 6 is connected to 5 but not 2, it separates 5 from 2 (partition \ref{eqn:bot3}). The bottom partition is now equitable and isomorphic to the top.


After the refinement procedure ends, \saucy{} builds isomorphic OPP (\ref{eqn:OPP3}), and starts exploring it by mapping 4 to 14, 12, 13, and 15. However, this phase of the search is superfluous, since we know that OPP (\ref{eqn:OPP3}) violates the graph's edge relation, and its further exploration will always result in  conflicts. Another case of a conflicting isomorphic OPP is when two corresponding \emph{singleton cells} of the top and bottom partitions have different connections to the other \emph{singleton cells} of their own partition. In this case, the conflict is again overlooked by \saucy{}'s conventional refinement procedure, since singleton cells cannot be partitioned to smaller cells (i.e., no new cell splitting occurs), and hence, the top and bottom partitions remain isomorphic after this step of refinement.

\begin{figure}[t]
\centering
\includegraphics[scale=0.5,keepaspectratio=true]{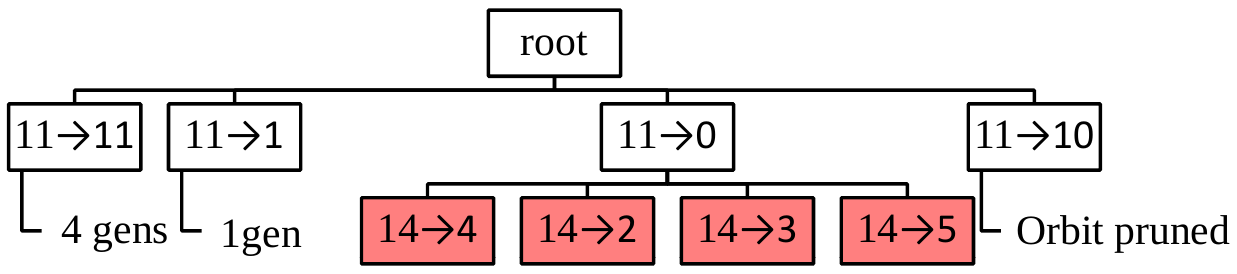}
\caption{The search tree constructed by \saucy{} 3.0 for the graph in Figure \ref{fig:graph}.}
\label{fig:tree2}
\end{figure}

To detect the conflicts that might remain undetected during partition refinement, we enhanced \saucy{}'s partition refinement in two ways; 1) the isomorphism of the bottom partition to the top is checked \emph{after each refinement step, rather than after each time a new split occurs}, and 2) in addition to the isomorphism check, we also ensure that \emph{the connections of each newly created cell on the bottom match the connections of its corresponding cell on the top}. These two new checks verify that the top and bottom partitions remain isomorphic and conforming (according to the graph's edge relation) after each refinement step. In our implementation, the overhead of the first check is negligible, as it is performed within the main refinement loop, but the second check requires an extra iteration over the outgoing edges of the vertices of the newly created cells. We would like to emphasize that our enhancement is \emph{enabled} by the OPP-encoding of permutations that is unique to \saucy{}'s search for automorphisms.

Figure \ref{fig:tree2} shows the search tree for the graph in Figure \ref{fig:graph} when our new simultaneous refinement is invoked. Comparing this search tree to that in Figure \ref{fig:tree1}, the number of conflicts is reduced from 16 to 4.

\section{The Validity of Matching OPP Pruning}
\label{sec:matching}

When matching OPP $\pi$ is encountered in the search, \saucy{} ``constructs'' a permutation $\alpha$ from $\pi$ by mapping the vertices in matching cells identically. It then uses $\alpha$ to prune the entire subtree rooted at this OPP in one of  two ways; either 1) $\alpha$ is an automorphism of the graph, which means that the subtree is a coset of the stabilizer subgroup, and $\alpha$ is a coset representative, or 2) $\alpha$ is not an automorphism, which indicates that the subtree is not a coset, and the search for a coset representative in that subtree will always fail. In this section, we show that, if $\pi$ is found to be matching by our enhanced simultaneous refinement (described in Section \ref{sec:refine}), the second case cannot occur, i.e., $\alpha$ must always be an automorphism of the graph. The proof of this claim is presented next.

Assume that $\pi$ is an OPP that is found matching by our enhanced refinement procedure. This means that $\pi$ is equitable, isomorphic, matching, and conforming according to $G$'s edge relation. Let $\alpha$ be the permutation that corresponds to $\pi$, i.e., the permutation that maps the vertices in $\pi$'s non-singleton cells identically. To show by contradiction that $\alpha$ is a symmetry of $G$, assume that it is not. Then, there must be an edge in $G^\alpha$ that does not exist in $G$ (or vice versa). Assume that this edge connects $v_1$ to $v_2$. Trivially, both $v_1$ and $v_2$ cannot be mapped identically in $\alpha$, otherwise, an edge between $v_1$ and $v_2$ in $G$ would map to the exact same edge in $G^\alpha$. Hence, permutation $\alpha$ either maps $v_1$ to $v'_1$ ($v_1\ne v'_1$), or $v_2$ to $v'_2$ ($v_2\ne v'_2$), or both. We first consider the case where $v_1$ is mapped to $v'_1$ but $v_2$ is mapped identically (this is similar to the case where $v_2$ is mapped to $v'_2$ but $v_1$ is mapped identically). This case contradicts our assumption that $\pi$ is equitable, since $v_1$ and $v'_1$ were both singleton cells of $\pi$, and having an edge between $v_1$ and $v_2$ but not between $v'_1$ and $v_2$ would imply further refinement on $\pi$. Now consider the case where $v_1$ is mapped to $v'_1$ and $v_2$ to $v'_2$. This case contradicts our assumption that $\pi$ is conforming according to $G$'s edge relation, since $v_1$, $v_2$, $v'_1$ and $v'_2$ were all singleton cells of $\pi$, and having an edge between $v_1$ and $v_2$ but not between $v'_1$ and $v'_2$ would violate $G$'s edge relation.

\section{Experimental Evaluation}
\label{sec:results}

We implemented our simultaneous partition refinement technique in \saucy{} 3.0, and tested its performance on 1445 graph benchmarks drawn from a wide variety of domains. Our experiments were conducted on a SUN workstation equipped with a 3GHz Intel Dual-Core CPU, a 6MB cache and an 8GB RAM, running the 64-bit version of Redhat Linux. A time-out of 1000 seconds was applied. Table \ref{tab:benchmarks} lists the benchmark families used in our experiments. For these families, the name, the number of instances, the size of the smallest and largest instances, and a short description are provided. The families are divided into four categories. These categories were chosen based on the general construction of the graphs, considering metrics such as the number of vertices and edges, connectivity and sparsity. The first category is the Miyazaki graphs \cite{Miyazaki_1997,miyazaki-graphs}, which \nauty{} takes exponential time to process. The second category contains benchmarks used to test earlier versions of \saucy{}. It represents graphs from various domains, such as logic circuits and their physical layouts \cite{Velev-DAC02,ISPD2005}, internet routers \cite{Cheswick00,Govindan00}, and road networks in the US states and its territories \cite{USCensusBureau}. The third category includes CNF benchmarks from the international SAT 2011 competition \cite{SATCompetition}. The fourth category consists of graphs not previously reported in graph automorphism or satisfiability research. These graphs were proposed for testing community-detection algorithms \cite{Lancichinetti-2009,binnet} \footnote{We used the implementation of the algorithm described in \cite{Lancichinetti-2009} (available at \cite{binnet}) to generate 33 undirected and unweighted binary networks. We set the number of nodes to $\{1,...,9\}\times \{10^3, 10^4, 10^5\}$ and $\{1,...,6\}\times 10^6$ (generating larger networks required more than 8GB RAM), and fixed the remaining parameters in all instances. Specifically, we set the average degree to 2, the max degree to 4, the mixing parameter to 0.1, the minimum community size to 20, and the maximum community size to 50.}.
\begin{table}[t]
  \centering
  \caption{Benchmark families}
  \resizebox{\textwidth}{!} {
    \begin{tabular}{p{2.3cm}|C{1.4cm}|C{1.1cm}C{1.15cm}|C{1.5cm}C{1.6cm}|p{3.6cm}}
    \hline\noalign{\smallskip}
    Family & Instances & \multicolumn{2}{c|}{Smallest Instance} & \multicolumn{2}{c|}{Largest Instance} & Description \\
           &           &         vertices       &     edges     &         vertices     &       edges    &             \\
    \hline\noalign{\smallskip}
    {\tt mz} \cite{Miyazaki_1997,miyazaki-graphs}  & 25 & 40 & 60  & 1,000 & 1,500 & Original Miyazaki graphs \\
    {\tt cmz} \cite{miyazaki-graphs}     & 46 & 120 & 90 & 200 & 1,900 & ({\tt mz}), and their variants \\
    {\tt mz-aug} \cite{miyazaki-graphs}  & 25 & 40 & 92  & 1,000 & 2,300 &  designed to mislead the\\
    {\tt mz-aug2} \cite{miyazaki-graphs} & 24 & 96 & 152 & 1,200 & 1,900 &  \bliss{} cell selector \\
    \hline\noalign{\smallskip}
    {\tt circuit} \cite{Velev-DAC02,ISPD2005} & 33 & 3,575   & 14,625  & 4,406,950  & 8,731,076 & \saucy{} benchmarks from\\
    {\tt router} \cite{Cheswick00,Govindan00} & 3 & 112,969  & 181,639   & 284,805 & 428,624  & place-route, verification, \\
    {\tt roadnet} \cite{USCensusBureau} & 56 & 1,158   & 1,008 & 1,679,418 & 2,073,394    & routers \& road networks \\
    \hline\noalign{\smallskip}    
    {\tt application} \cite{SATCompetition} & 300 & 464 & 2,066 & 32,813,545 & 65,487,132 & SAT 2011 application,\\
    {\tt crafted} \cite{SATCompetition} & 300 & 105 & 320 & 776,820 & 3,575,337 & crafted and random\\
    {\tt random} \cite{SATCompetition}  & 600 & 1,165 & 5,375 & 310,000 & 680,000 & CNF instances\\
    \hline\noalign{\smallskip}
    {\tt binnet} \cite{Lancichinetti-2009,binnet} & 33  & 1,000 & 720 & 6,000,000  & 4,391,515  & binary networks \\    
    \hline\noalign{\smallskip}
    \end{tabular}%
  }
  \label{tab:benchmarks}
  \vspace{-10pt}
\end{table}%

\begin{figure}[t]
\centering
\includegraphics[scale=0.55,keepaspectratio=true]{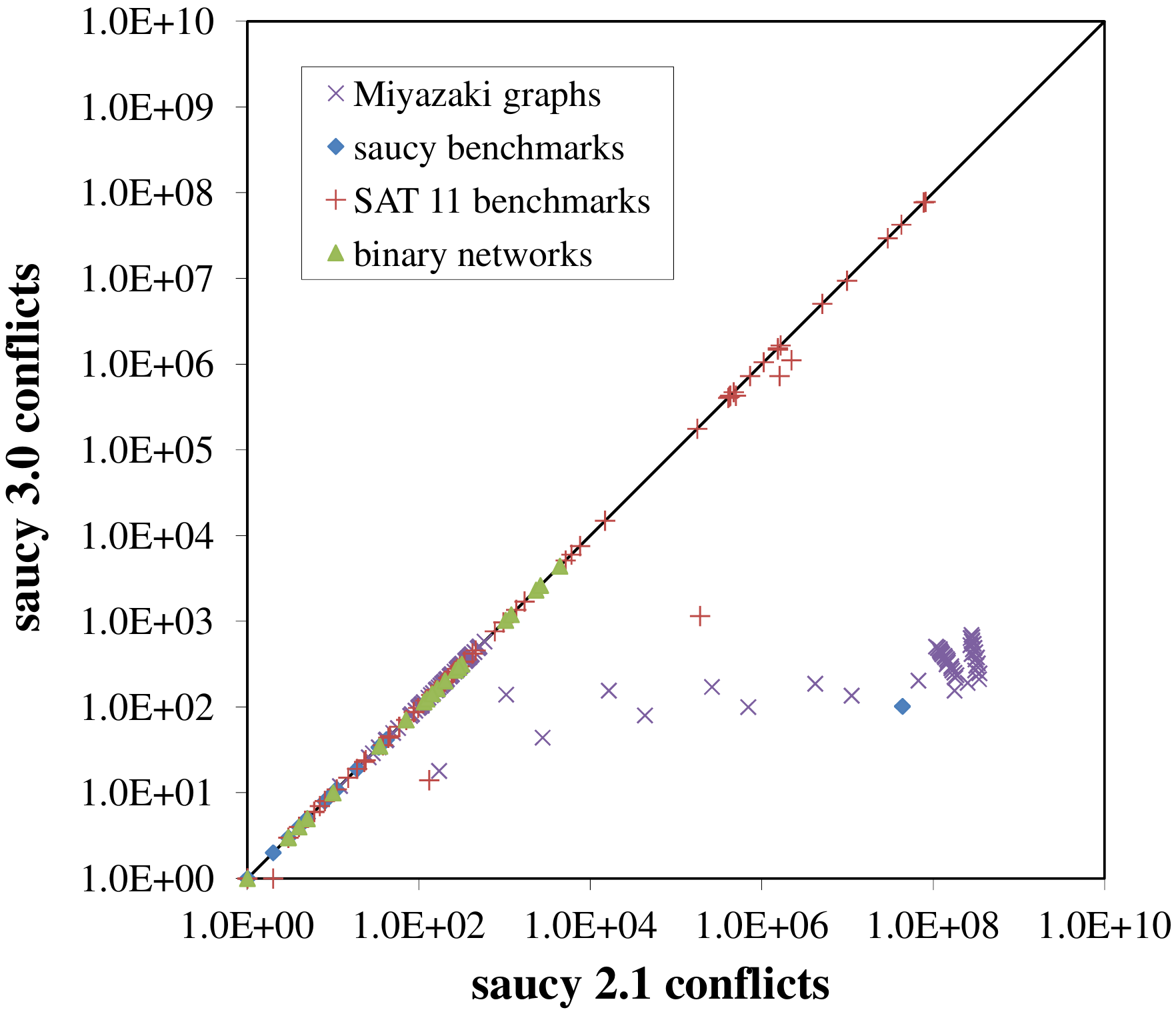}
\caption{Number of conflicts returned by \saucy{} 3.0 versus \saucy{} 2.1.}
\label{fig:conflicts}
\end{figure}

Figure \ref{fig:conflicts} compares the number of conflicts produced by \saucy{} 3.0 and \saucy{} 2.1. If a benchmark is not processed within the time-out, the number of conflicts encountered right before termination is reported. The results show that \saucy{} 3.0 always produces fewer or the same number of conflicts. This is expected, as our proposed refinement procedure anticipates and avoids certain conflicts that might arise in \saucy{} 2.1. Of all the benchmark families, {\tt mz-aug} and {\tt mz-aug2} benefit most from the new refinement procedure. For these two families, the highest number of conflicts reported by \saucy{} 3.0 was 696 (for {\tt mz-aug-50}). In contrast, the number of conflicts reported by \saucy{} 2.1 was at least 10,000 for 46 out of 49 {\tt mz-aug} and {\tt mz-aug2} instances. Of the remaining two Miyazaki families, {\tt mz} did not experience any change in its number of conflicts, and {\tt cmz} showed a slight improvement for 5 out of its 46 instances (8 fewer conflicts were reported for those 5 instances). Of the graphs from circuits, internet routers, and road networks, only one instance (from {\tt circuit}) showed significant conflict reduction (from 43 million to only 102). The remaining instances produced the same number of conflicts (not more than 42) in \saucy{} 3.0 and \saucy{} 2.1. Of the 1200 CNF benchmarks, only 72 (15 from application and 57 from crafted) encountered conflicts in \saucy{} 2.1, and only 12 (all from crafted) experienced a reduction in the number of conflicts. The smallest reduction was 1 and the largest was 2.9 million. The {\tt binnet} instances also produced the same results in both versions of \saucy{}. The reported number of conflicts for those instances ranged from no conflicts to 4,412. 

\begin{figure}[t]
\centering
\includegraphics[scale=0.75,keepaspectratio=true]{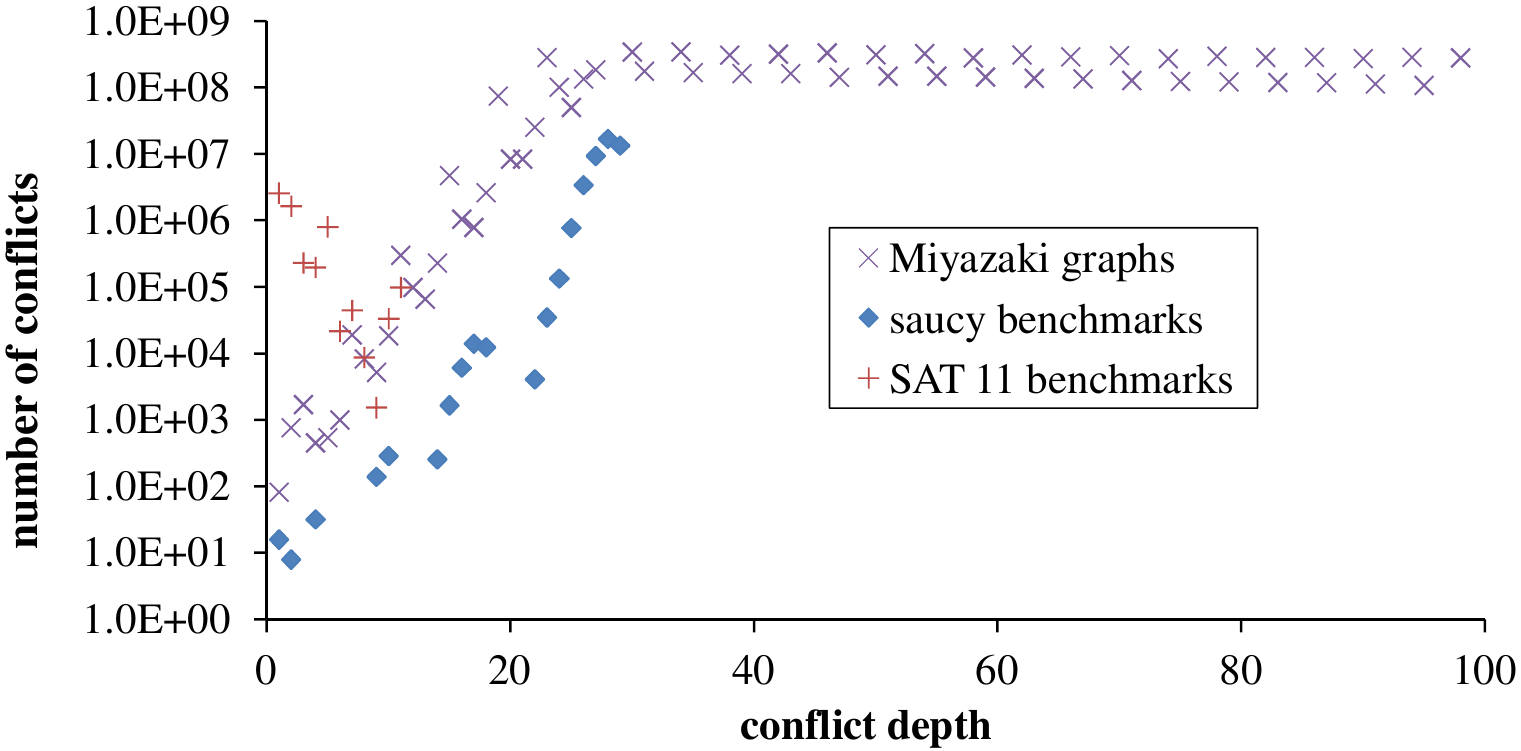}
\caption{Histogram of the conflict depths captured by \saucy{} 3.0.}
\label{fig:histogram}
\end{figure}

Figure \ref{fig:histogram} shows the distribution of \emph{depth} of the conflicts that were captured and avoided by \saucy{} 3.0. Recall that the new refinement procedure in \saucy{} 3.0 prunes some subtrees that are explored by \saucy{} 2.1. Suppose that one such subtree is found to be conflicting at level $l$ in \saucy{} 3.0, but leads to $c$ conflicts in \saucy{} 2.1, where the $n$-th conflict ($1\le n\le c$) occurs at level $l_n$. Trivially, $l_n\ge l$. We define the depth of the $n$-th conflict as $d=l_n - l$. If $d=0$, both \saucy{} 3.0 and \saucy{} 2.1 capture the conflict at the same time. If $d>0$, \saucy{} 3.0 anticipates and avoids the conflict $d$ levels sooner than it occurs in \saucy{} 2.1. We use conflict depth as a numeric criterion to evaluate the effectiveness of our new refinement procedure. The results in Figure \ref{fig:histogram} show that the deepest conflicts captured by \saucy{} 3.0 occur in the instances of Miyazaki families. The greatest reported depth was 98, which occurred $2.8\times 10^8$ times for {\tt mz-aug-50}. The only benchmark from the {\tt circuit} family that had significant conflict reduction produced conflict depth of up to 29, where the largest conflict depth happened $1.3\times 10^7$ times. For the CNF benchmarks, the deepest reported conflict had a depth of 11, and occurred roughly $10^5$ times. The histogram in Figure \ref{fig:histogram} excludes the results for binary networks, since all those conflicts were reported at depth 0.

\begin{figure*}[p]
\centering
\includegraphics[scale=0.55,keepaspectratio=true]{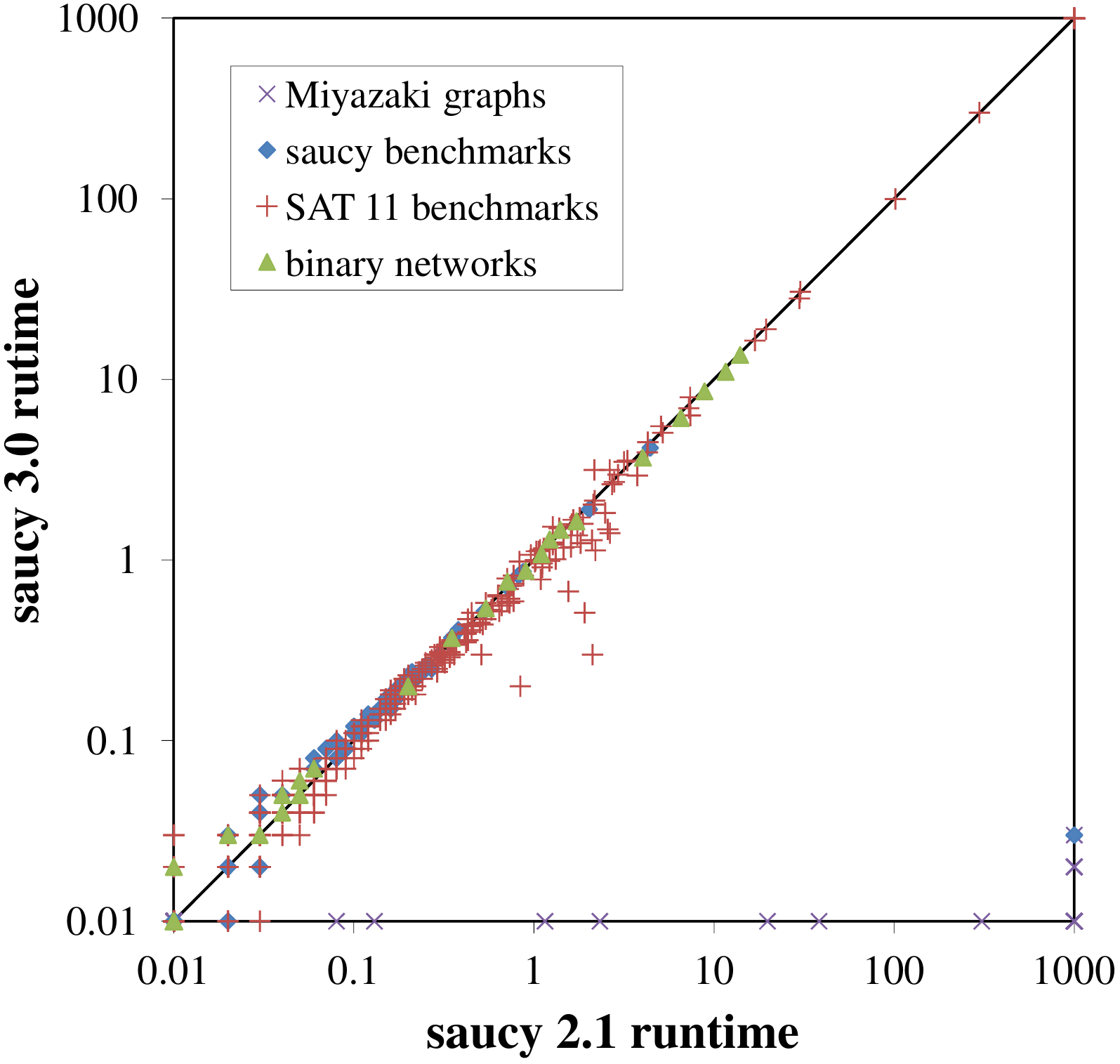}
\caption{Runtime of \saucy{} 3.0 versus \saucy{} 2.1 (timeout is 1000 seconds).}
\label{fig:runtimes}
\vspace{20pt}
\includegraphics[scale=0.55,keepaspectratio=true]{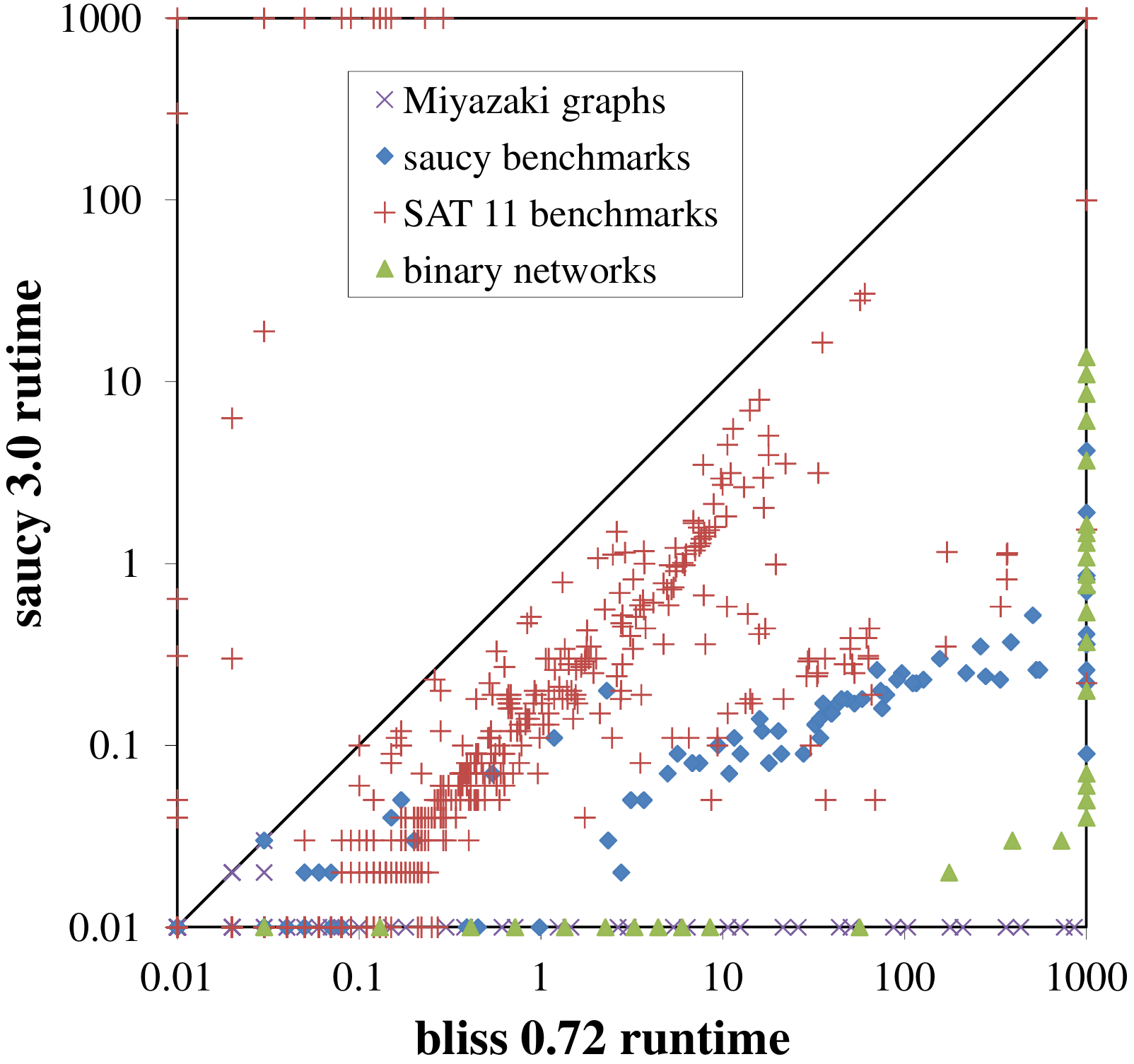}
\caption{Runtime of \saucy{} 3.0 versus \bliss{} 0.72 (timeout is 1000 seconds).}
\label{fig:sbruntimes}
\end{figure*}

The runtime comparison between \saucy{} 3.0 and \saucy{} 2.1 is depicted in Figure \ref{fig:runtimes}. For the families of {\tt mz-aug} and {\tt mz-aug2}, we observed an exponential speedup when our proposed refinement procedure was invoked. Of the 49 instances in these two families, \saucy{} 3.0 solved all in less than a second, while \saucy{} 2.1 failed to process 39 within the time-out limit.
For the {\tt mz} and {\tt cmz} families, \saucy{} 2.1 and 3.0 had comparable runtimes. The instances of {\tt router}, {\tt roadnet}, and {\tt binnet} did not experience much change either. For the {\tt circuit} family, the results were comparable, except for one benchmark that was solved by \saucy{} 3.0 in a second but remained unsolved in \saucy{} 2.1. Interestingly enough, we did not observe any major improvement in the runtimes of the SAT 11 CNF benchmarks, although conflict reduction of up to 2.9 million was reported for some of those instances. Our further analysis revealed that high reduction in the number of conflicts was reported for instances that timed out in both \saucy{} 3.0 and \saucy{} 2.1, and the reduction in the remaining instances was not significant enough to reflect a major improvement in runtimes. Note that the runtimes reported in Figure \ref{fig:runtimes} match with the number of conflicts reported in Figure \ref{fig:conflicts}. In fact, fewer conflicts generally led to better runtimes.

In order to evaluate the performance of \saucy{} 3.0 versus state-of-the-art graph automorphism tools, we ran \bliss{} (version 0.72, available at \cite{bliss-0.72}) on all the 1445 benchmarks listed in Table \ref{tab:benchmarks}, and compared its runtimes to those obtained from \saucy{} 3.0. This comparison is shown in Figure \ref{fig:sbruntimes}. Of the four Miyazaki graph families, \bliss{} showed difficulties in processing the instances of {\tt cmz} (took up to 856 seconds to complete all those instances), but processed the remaining three families in less than a second. 
In contrast, \saucy{} solved all Miyazaki graphs in less than a second. Furthermore, \bliss{} timed out on 8 and 3 out of 33 and 56 instances of the {\tt circuit} and {\tt roadnet} families, respectively, but solved the remaining instances of those two families and all 3 instances of {\tt router} in 550 seconds. This was while \saucy{} solved all the 92 instances of these three families in 5 seconds (processed 90 in less than a second). For the CNF benchmarks, \saucy{} and \bliss{} showed mixed results. Of the 600 crafted and application instances, \bliss{} failed to process 4 crafted and 3 application instances, whereas, \saucy{} failed to process 17 crafted instances, but solved all application instances. The 4 crafted benchmarks that were unsolved by \bliss{} were also unsolved by \saucy{}. This means that \bliss{} solved 13 crafted instances that \saucy{} failed to process, and \saucy{} solved 3 application instances that \bliss{} did not solve. Of the remaining crafted and application benchmarks, \bliss{} solved 541 
in less than 10 seconds, and 52 
in 366 seconds, while \saucy{} solved 577 
in less than 10 seconds, and 6 
in 300 seconds. Both \saucy{} and \bliss{} solved all random benchmarks in less than a second. Overall, the results in Figure \ref{fig:sbruntimes} indicate that \saucy{} outperformed \bliss{} on the majority of SAT 11 benchmarks. For binary networks, \saucy{} consistently produced better results. Specifically, \saucy{} solved all 33 instances of {\tt binnet} in 14 seconds (the largest runtime was 13.67 seconds which was reported for the largest instance of this family with $6\times 10^6$ vertices), but \bliss{} timed out on 19, and solved the remaining in 727 seconds. 
\begin{figure}[t]
\centering
\includegraphics[scale=0.75,keepaspectratio=true]{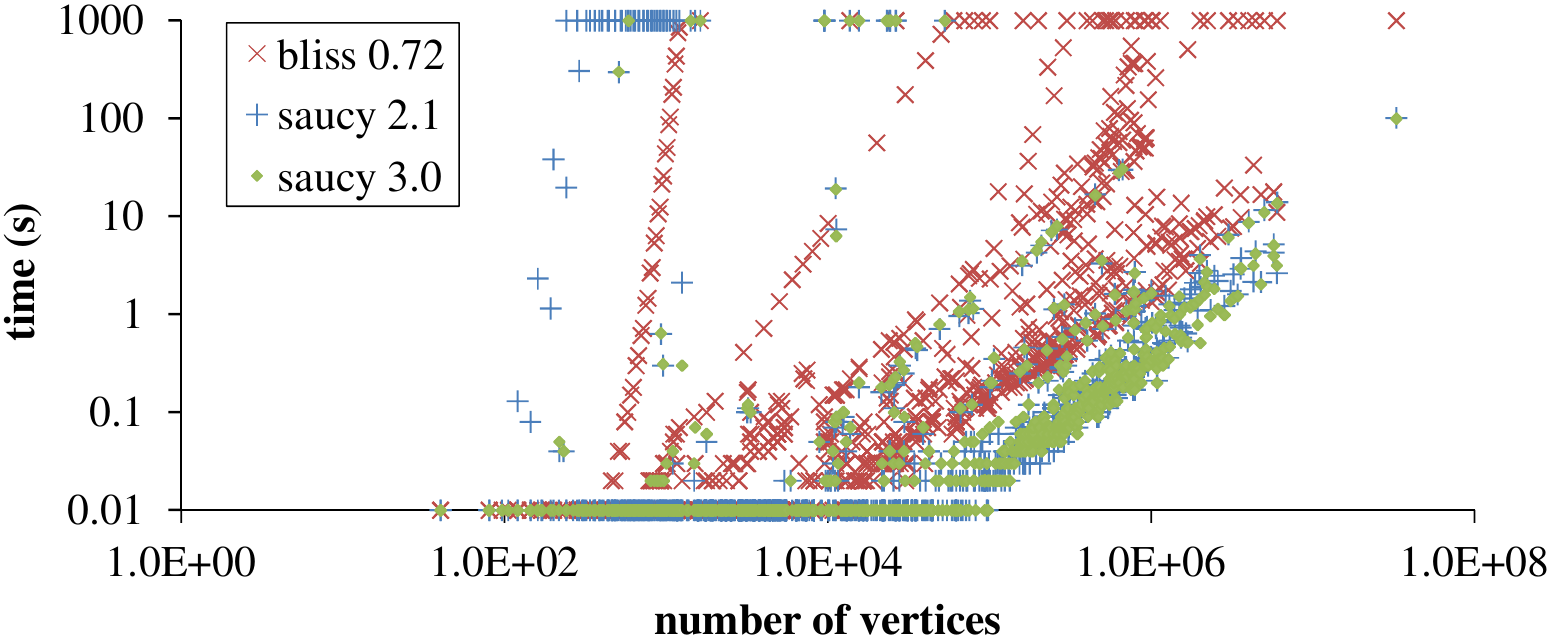}
\caption{Runtimes of \saucy{} 3.0, \saucy{} 2.1, and \bliss{} 0.72 as a function of graph size.}
\label{fig:time_vs_size}
\end{figure}

As part of our study, we also ran \nishe{} 0.1 \cite{nishe-0.1} on all the graph benchmarks in our suite, and compared its results to \saucy{} 3.0. In general, we observed that the runtimes of \nishe{} and \saucy{} were comparable for the Miyazaki graphs. For the remaining benchmarks, however, \nishe{} exhibited poor performance compared to \saucy{} and \bliss{}. In particular, it failed to process (either timed out or had a segmentation fault) 59 out of 92 \saucy{} benchmarks, 950 out of 1200 CNF instances, and 24 out of 33 binary networks.

Figure \ref{fig:time_vs_size} shows the runtimes of \saucy{} 3.0, \saucy{} 2.1, and \bliss{} 0.72 as a function of graph size for all the 1445 benchmarks listed in Table \ref{tab:benchmarks}. As this figure suggests, the smaller instances seem to be more challenging for \saucy{}. This is particularly not true of \bliss{}, as \bliss{} tends to produce larger runtimes for larger instances. The smallest instance that \saucy{} 3.0 timed out on had 583 vertices, and the largest had 52,786 vertices, while these numbers were respectively reported to be 1,620 and 33 million for \bliss{} 0.72. Of the 446 benchmarks with more than 52,786 vertices, \saucy{} 3.0 solved 389 in less than a second, and processed the rest in 100 seconds, while \bliss{} 0.72 solved 213 in less than a second, took up to 550 seconds to process 200, and timed out on 33. On the other hand, of the 999 benchmarks that had less than 52,786 vertices, \saucy{} 3.0 solved 979 in less than a second, timed out on 17, and took up to 550 seconds to process the rest, whereas, \bliss{} 0.72 processed 946 in less than a second, timed out on 4, and processed the remaining in 856 seconds. To investigate the reason why \saucy{} 3.0 did not perform as expected on relatively small instances, we examined the effect of different decision heuristics on the 17 benchmarks that \saucy{} failed to process. Interestingly, 4 out of those 17 benchmarks were solved in less than a second with an alternative decision heuristic. Of those 4, one was reported to be unsolved by \bliss{} 0.72. These results suggest that branching decisions play a crucial role in minimizing the time for automorphism search. We plan to pursue the effect of decision heuristics in our future research.

\section{Conclusions}
\label{sec:conc}

In this work, we have advanced the state of the art in algorithms for solving graph automorphism, which finds applications in many fields. Our technique takes advantage of a unique feature in the \saucy{} algorithm --- the representation of partial permutations (search nodes) in terms of ordered partition pairs. Previously, these partitions were refined one at a time, but we have now developed {\em simultaneous partition refinement}, which  allows \saucy{} to anticipate possible future conflicts and prune the search tree early. This optimization significantly improves runtime on several benchmark families, including the ones suggested by Miyazaki \cite{Miyazaki_1997} for further study because \nauty{} provably requires exponential time on these benchmarks. Our empirical comparisons show that our implementation \saucy{} 3.0 outperforms the competition on most available benchmarks. Our ongoing work is focused on several benchmarks where \saucy{} 3.0 is outperformed by \bliss{} 0.72. Preliminary analysis suggests that these benchmarks tend   to be small, which may be due to subtle inefficiencies in our implementation rather than asymptotic bottlenecks. We hope that our future research will shed additional light on this.

\bibliographystyle{plain}
\addcontentsline{toc}{section}{References}
\bibliography{ConflictAnticipation}

\end{document}